\documentclass[aps,pre,showpacs,showkeys]{revtex4}

\usepackage{amsmath}
\usepackage{amssymb}
\usepackage{graphicx}
\usepackage{psfrag}

\newcommand{\mean}[1]{\left\langle #1 \right\rangle}
\newcommand{\smean}[1]{\langle #1 \rangle}
\newcommand{\skap}[2]{\left\langle #1 , #2 \right\rangle}

\begin{document}

\title{Subthreshold dynamics of the neural membrane potential\\driven by stochastic synaptic input}
\author{Ulrich Hillenbrand}
\email[]{Ulrich.Hillenbrand@dlr.de}
\affiliation{Institute of Robotics and Mechatronics\\German Aerospace Center\\Oberpfaffenhofen, 82234 Wessling, Germany}

\begin{abstract}
In the cerebral cortex, neurons are subject to a continuous bombardment of synaptic inputs originating from the network's background activity. This leads to ongoing, mostly subthreshold membrane dynamics that depends on the statistics of the background activity and of the synapses made on a neuron. Subthreshold membrane polarization is, in turn, a potent modulator of neural responses. The present paper analyzes the subthreshold dynamics of the neural membrane potential driven by synaptic inputs of stationary statistics. Synaptic inputs are considered in linear interaction. The analysis identifies regimes of input statistics which give rise to stationary, fluctuating, oscillatory, and unstable dynamics. In particular, I show that (i) mere noise inputs can drive the membrane potential into sustained, quasiperiodic oscillations (noise-driven oscillations), in the absence of a stimulus-derived, intraneural, or network pacemaker; (ii) adding hyperpolarizing to depolarizing synaptic input can increase neural activity (hyperpolarization-induced activity), in the absence of hyperpolarization-activated currents.
\end{abstract}


\maketitle

\noindent Published as {\em Physical Review E} {\bf 66}, 021909 (2002).

\section{Introduction}

Cortical pyramidal cells fire action potentials at an average spontaneous rate of about 10 spikes$/$s in waking animals \cite{Steriade_etal74,Steriade78}. At such a low spike rate, it is clear that most cortical neurons spend a significant amount of time with their membrane potential well below the threshold for spike activation. On the other hand, a cortical pyramidal cell receives roughly 10000 synapses \cite{Defelipe&Farinas92}, mostly from other cortical neurons. Since individual postsynaptic events cause transient increases in membrane conductance, it follows that the dynamics of membrane potentials is largely controlled by subthreshold stimulation from the continuous network activity. Subthreshold membrane polarization is, in turn, a potent modulator of stimulus-driven spike activity \cite{Pare_etal98,Azouz&Gray99}.

In this paper, I analyze the subthreshold dynamics of the membrane potential driven by stochastic synaptic activity of general stationary statistics. Such conditions are given in neurons that do not respond to an external stimulus, but are exposed to the network's spontaneous or stimulus-driven background activity. The generation of postsynaptic potentials (PSPs) and their propagation along the dendrites of a neuron are modeled in a rather simple way to allow for a thorough analytical treatment. Accordingly, the focus is on generic patterns of behavior rather than on quantitative results. Some of the conclusions are discussed in relation to the experimental literature.

\section{Modeling synaptic responses}

The potential $V$ across a local patch of passive membrane is described by
\begin{equation}
\label{Vni}
\frac{d}{dt} V = -\frac{1}{\tau_m} \, V + \frac{1}{\tau_m \, g_m} \, I \ ,
\end{equation}
where $\tau_m$ and $g_m$ are the passive membrane time constant and leak conductance, respectively, and $I$ is the current passed along the dendrites from other parts of the cell. The membrane's resting potential is set to zero. After a synaptic input has been received on the considered patch of membrane, the potential obeys
\begin{equation}
\label{Vi}
\frac{d}{dt} V = -\frac{1}{\tau_m} \, V + \frac{1}{\tau_m \, g_m} \, I + \frac{g_s}{\tau_m \, g_m} \, \left( V_s - V \right) \ ,
\end{equation}
where $V_s$ and $g_s$ are the synaptic reversal potential and conductance, respectively. Let $V_0(t)$ and $V_{\rm in}(t)$ be solutions to Eqs.\ (\ref{Vni}) and (\ref{Vi}), respectively, with $V_0(0) = V_{\rm in}(0) = V(0)$. Synaptic ion channels are open for a brief period $\delta_s \ll \tau_m$ \footnote{Conduction times of ionotropic synapses are roughly 1 ms, while the membrane time constant is usually several 10 ms \cite{Hille92}. The other type of synapse, the metabotropic synapse, modulates the membrane potential on a much longer time scale of seconds \cite{Hille92} and is not considered here.}. At time $t = \delta_s$, when synaptic channels close, the deflection of the membrane potential due to the synaptic input is
\begin{equation}
\label{syn_deflect}
V_{\rm in}(\delta_s) - V_0(\delta_s) = \frac{\delta_s}{\tau_m} \, \frac{g_s}{g_m} \, \left[ V_s - V(0) \right] + O\!\left[\left(\frac{\delta_s}{\tau_m}\right)^2\right] \ .
\end{equation}
This deflection propagates along the cell's dendrites. Far away from its point of origin, I model the synaptic response as a PSP. In a passive cable, the rise time and amplitude of a PSP depend on the time course of the synaptic current, and the relative locations of the synapse and the point on the membrane at which the PSP is observed; the decay-time constant approaches $\tau_m$ for long times \cite{Rall62,Rall70,Rall77,Tuckwell88a}. However, computer-simulation studies involving realistic cell morphologies \cite{Chitwood_etal99,Jaffe&Carnevale99} and voltage-dependent dendritic conductances \cite{Destexhe&Pare99} have revealed that PSPs in real neurons may be less variable than suggested by a cylindrical passive-cable model. A coarse but, for the present analysis,  sufficient approximation to a PSP is given by the impulse response of a second-order low-pass filter,
\begin{equation}
\label{PSP}
\Lambda\!\left(\gamma,V_s,t_0;t\right) :=
\gamma \, \left[ V_s - V(t_0) \right] \, \frac{t - t_0}{\tau} \, \exp\!\left( 1 - \frac{t - t_0}{\tau} \right) \, \Theta\!\left(t - t_0\right) \ ,
\end{equation}
with the unit-step function
\begin{equation}
\Theta(t) :=
\left\{ \begin{array}{ll}
0 & \mbox{for $t \le 0$,} \\
1 & \mbox{for $t > 0$.}
\end{array} \right.
\end{equation}
The PSP's amplitude is $\gamma \, \left[ V_s - V(t_0) \right]$, with the factor
\begin{equation}
\label{gamma}
\gamma := a \, \frac{\delta_s}{\tau_m} \, \frac{g_s}{g_m} > 0 \ .
\end{equation}
Thus, the PSP is initiated at time $t_0$, has a rise time and decay-time constant $\tau$, is attenuated or amplified by a factor $a$ [cf.\ Eq.\ (\ref{syn_deflect})], and is assumed to propagate instantaneously. It qualitatively captures the basic properties of real PSPs of having a finite rise time and an exponential decay phase. It is chosen here for its convenience for analysis.

Postsynaptic conductance changes are very local compared to the extended dendritic trees on which synapses make contacts. It is therefore a reasonable approximation to treat them as noninteracting. The total membrane potential under synaptic control is hence given by the sum
\begin{equation}
\label{V_tot}
V(t) = \sum_{i=1}^{\infty} \Lambda\!\left(\gamma_i,s_i,t_i;t\right)
\end{equation}
for the whole cell. Here $t_1 \le t_2 \le \ldots$ are the times of synaptic input received by a neuron; $\gamma_i$ and $s_i$ are the amplitude-related factor defined in Eq.\ (\ref{gamma}) and the reversal potential of the $i$th synaptic input, respectively. In Sec.\ \ref{delays}, I will address effects of delays in the propagation of PSPs.

\section{Analysis and results}

Upon inspection of Eqs.\ (\ref{PSP}) and (\ref{V_tot}), it is clear that there is an equivalence relation between the statistics of the $\gamma_i$ and of the pairs $(s_i,t_i)$. Higher values of $\gamma_i$ have the same effect on the dynamics of $V(t)$ as shorter intervals $t_{i+1} - t_i$ between successive stimuli with $s_i = s_{i+1}$. In order to simplify the analysis, without limiting the dynamic repertoire of $V(t)$, it is preferable to restrict to one value $\gamma \equiv \gamma_i$. In this section, I shall thus derive analytical results on the dynamics
\begin{equation}
\label{V_dyn}
V(t) = \gamma \, \sum_{i=1}^{\infty} \left[ s_i - V(t_i) \right] \, \frac{t - t_i}{\tau} \, \exp\!\left( 1 - \frac{t - t_i}{\tau} \right) \, \Theta\!\left(t - t_i\right) \ .
\end{equation}
Moreover, the results will be illustrated by computer simulations where appropriate.

Arguably, the ``obvious'' approach to the problem is to specify the distribution functions for the point process that models the times $t_i$ of stimulus events and write down integral equations for the moments of $V(t)$. However, we shall take a different approach. We will start by casting the dynamics in the form of a Markov chain. There are two significant advantages proceeding this way. First, it will allow us to go quite far with the analysis without being specific about the stimulus process. Only at some later point will it be profitable to specify the statistics of stimulus times. Second, making use of the Markov property, we will gain insight not only into the dynamics of moments of the membrane potential, but also into the temporal pattern of {\em individual} trajectories $V(t)$.

\subsection{Markov formulation of the dynamics of the membrane potential\label{Markov_form}}

Introducing the notation
\begin{eqnarray}
x_j & := & V(t_j) = \gamma \sum_{i = 1}^{j - 1} \left[s_i - V(t_i)\right] \, \frac{t_j - t_i}{\tau} \, \exp\!\left( 1 - \frac{t_j - t_i}{\tau} \right) \ ,\\
y_j & := & \gamma \sum_{i = 1}^{j - 1} \left[s_i - V(t_i)\right] \,
\exp\!\left( -\frac{t_j - t_i}{\tau} \right) \ ,\\
r_j & := & t_{j + 1} - t_j \ ,
\end{eqnarray}
we can reformulate the dynamics of Eq.\ (\ref{V_dyn}) for the discrete times $t = t_j$ as an iteration of a combination of two stochastic maps ${\cal R}(r)$ and ${\cal S}(s)$,
\begin{equation}
\label{map_dyn}
{x_j \choose y_j} = {\cal R}(r_{j - 1}) \circ {\cal S}(s_{j - 1}) \,
{x_{j - 1} \choose y_{j - 1}} \ , \quad x_1 = y_1 = 0 \ ,
\end{equation}
\begin{eqnarray}
{\cal S}(s): && {x \choose y} \mapsto {x
\choose y + \gamma \, (s - x)} \ ,\\
{\cal R}(r): && {x \choose y} \mapsto {\left( x + e y \frac{r}{\tau}\right) e^{-r/\tau}
\choose y e^{-r/\tau}} \ .
\end{eqnarray}
The interstimulus times $r_j$ and the synaptic reversal potentials $s_j$ are stochastic variables, drawn independently from densities $u(r)$ on $\mathbb{R}_+$ and $v(s)$ on $\mathbb{R}$, respectively. These densities are determined by the neural network activity and the number and types of synapses on the neuron considered. Note that although there may well be statistical dependences between $r_j$ and $s_j$, and $(r_j,s_j)$ and $(r_{j'},s_{j'})$ ($j \not= j'$) as sampled at one {\em individual} synapse, these do not show up in the sequences $r_j$ and $s_j$ for {\em all} synaptic inputs to a cortical neuron.

In the present formulation of the dynamics, the synaptic input times $t_j$ are, like $x_j$ and $y_j$, stimulus-driven stochastic variables and may be incorporated by extending the system (\ref{map_dyn}) with the equation
\begin{equation}
\label{t_map}
t_j = t_{j - 1} + r_{j - 1} \ .
\end{equation}
This equation can be solved independently of Eq.\ (\ref{map_dyn}). In particular,
\begin{equation}
\mean{t_j} = (j - 1) \mean{r} + t_1 \ .
\end{equation}
Here and in the following, we encounter mean values of the types
\begin{equation}
\mean{f(s)} := \int_{-\infty}^{\infty} ds' \, v(s') \, f(s') \ ,\quad
\mean{f(r)} := \int_0^{\infty} dr' \, u(r') \, f(r') \ ,
\end{equation}
with $f$ being some function on the real numbers for which the integrals are defined.

The dynamics (\ref{map_dyn}) is a Markov chain. The transition probability corresponding to ${\cal S}(s)$ is
\begin{equation}
\label{pS_transit}
p_{\cal S}(x,y|x',y') = \int_{-\infty}^{\infty} ds \,
v(s) \, \delta(x - x') \, \delta\!\left[y - y' - \gamma (s - x')\right] \ ,
\end{equation}
and the one corresponding to ${\cal R}(r)$ is
\begin{equation}
\label{pR_transit}
p_{\cal R}(x,y|x',y') = \int_0^{\infty} dr \,
u(r) \, \delta\!\left[x - \left(x' + e y' \frac{r}{\tau}\right)
e^{-r/\tau}\right] \, \delta\!\left(y - y' e^{-r/\tau}\right) \ .
\end{equation}
Here $\delta$ is the Dirac delta function. Let $p(x,y)$ be a joint probability density for $x$ and $y$. Then
\begin{equation}
\mean{x^n y^m} := \int_{-\infty}^{\infty} dx'
\int_{-\infty}^{\infty} dy' \, p(x',y') \, x^{\prime n} y^{\prime m} \ ,\quad
n,m \in \mathbb{N} \ ,
\end{equation}
are the moments of $x$ and $y$. We want to know how the moments change under the action of ${\cal R}(r) \circ {\cal S}(s)$. For the action of ${\cal S}(s)$, we get
\begin{multline}
\label{momS_iter}
\mean{x^n y^m}_{\cal S} = \int_{-\infty}^{\infty} d\bar{x} \int_{-\infty}^{\infty} d\bar{y} \int_{-\infty}^{\infty} dx' \int_{-\infty}^{\infty} dy' \, p_{\cal S}(\bar{x},\bar{y}|x',y') \, p(x',y') \,
\bar{x}^n \bar{y}^m \\
= \sum_{h,i,j \in \mathbb{N} \atop h+i+j = m} {m \choose h,i,j} (-1)^i \gamma^{h+i} \mean{s^h} \mean{x^{n+i} y^j} \ ,
\end{multline}
with polynomial coefficients
\begin{equation}
{m \choose h,i,j} := \frac{m!}{h!\,i!\,j!} \ ,\quad
h+i+j = m \ .
\end{equation}
The action of ${\cal R}(r)$ yields
\begin{multline}
\label{momR_iter}
\mean{x^n y^m}_{\cal R} = \int_{-\infty}^{\infty} d\bar{x} \int_{-\infty}^{\infty} d\bar{y} \int_{-\infty}^{\infty} dx' \int_{-\infty}^{\infty} dy' \, p_{\cal R}(\bar{x},\bar{y}|x',y') \, p(x',y') \,
\bar{x}^n \bar{y}^m \\
= \sum_{k = 0}^n {n \choose k} \mean{\left(\frac{e r}{\tau}\right)^k e^{-(n+m) r/\tau}} \mean{x^{n-k} y^{m+k}} \ .
\end{multline}
Let $p_j(x,y)$ be the joint probability density of $x$ and $y$ at time $t_j$. By combining Eqs.\ (\ref{momS_iter}) and (\ref{momR_iter}), we can write down iteration equations for the moments,
\begin{equation}
\label{const_j_average}
\mean{x^n y^m}_j := \int_{-\infty}^{\infty} dx'
\int_{-\infty}^{\infty} dy' \, p_j(x',y') \, x^{\prime n} y^{\prime m} \ .
\end{equation}
The iterations can be solved successively for all $n$ and $m$, starting with the first moments. We shall solve for the first two moments, i.e., for $\smean{x}_j$, $\smean{y}_j$, $\smean{x^2}_j$, $\smean{xy}_j$, and $\smean{y^2}_j$. Note that the ensemble averages (\ref{const_j_average}) are not taken at constant time $t$, but rather at a constant number $j$ of synaptic inputs received, irrespective of the time $t_j$ of the $j$th input. As mentioned above, the times of synaptic inputs are additional random variables obeying Eq.\ (\ref{t_map}).

\subsection{Mean membrane potential\label{mean_V}}

\begin{figure}
\psfrag{e}{\raisebox{1mm}{\scriptsize $a_1$}}
\psfrag{gre}{\scriptsize $\;\gamma b_1$}
\centerline{\includegraphics[width=0.5\textwidth]{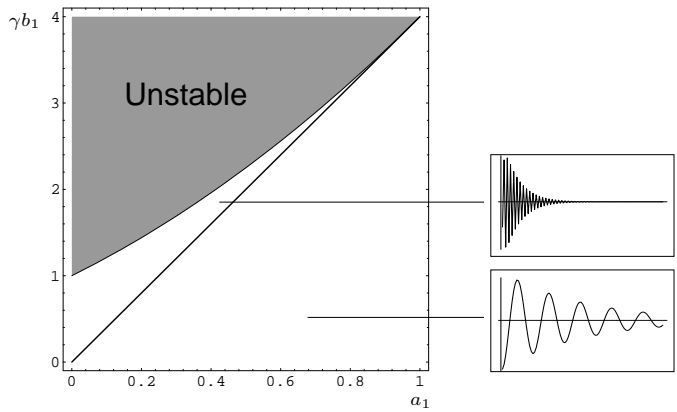}}
\caption[]{Space of stimulus parameters $a_1$ and $b_1$ that determine the dynamics of the mean membrane potential. The dynamics converges for $(a_1 + 1)^2 > \gamma b_1$. For $\gamma b_1 < 4 a_1$, the two eigenvalues given by Eq.\ (\ref{M_1_eigenvalues}) are complex conjugate. For $(a_1 + 1)^2 > \gamma b_1 > 4 a_1$, they are real and negative. The corresponding type of mean dynamics is depicted for these two regimes.
\label{a1_b1_space}}
\end{figure}

The iteration dynamics of the mean values obtained from Eqs.\ (\ref{momS_iter}) and (\ref{momR_iter}) is
\begin{equation}
\label{1.mom_iter}
{\mean{x}_j \choose \mean{y}_j} = \underbrace{\left(
\begin{array}{cc}
a_1 - \gamma b_1 & b_1 \\
-\gamma a_1 & a_1
\end{array}
\right)}_{=: M_1}
{\mean{x}_{j-1} \choose \mean{y}_{j-1}}
+ \gamma \mean{s} {b_1 \choose a_1} \ , \quad
\mean{x}_1 = \mean{y}_1 = 0 \ ,
\end{equation}
with the stimulus parameters $\smean{s}$ and
\begin{equation}
\label{stim_params_1}
\left.\begin{array}{rcl}
a_1 & := & \mean{e^{-r/\tau}} \\
b_1 & := & \mean{\frac{r}{\tau} e^{1 - r/\tau}}
\end{array}\right\} \in (0,1) \ .
\end{equation}
The dynamics of $\mean{x}_j$ and $\mean{y}_j$ depend on the eigenvalues of
$M_1$, and thus on the stimulus parameters $a_1$ and $b_1$. The eigenvalues
are
\begin{equation}
\label{M_1_eigenvalues}
\lambda_{1/2} := a_1 - \frac{\gamma b_1}{2}
\pm \frac{1}{2}\sqrt{\gamma^2 b_1^2
- 4 \gamma a_1 b_1} \ .
\end{equation}
For convergence of the dynamics, we require that
\begin{equation}
\left|\lambda_{1/2}\right| < 1 \quad \Longleftrightarrow \quad
\gamma b_1 < (a_1 + 1)^2 \ .
\end{equation}
Figure \ref{a1_b1_space} shows the parameter regions of convergence and divergence. In this parameter space, the vicinity of the point $a_1 = 1$, $b_1 = 0$ is occupied by high-frequency stimuli, i.e., with short interstimulus times $r$. A very low network activity, on the other hand, lies close to the point $a_1 = 0$, $b_1 = 0$. It turns out that for any input statistics, the mean of the membrane potential $V$ converges, if the factor $\gamma$, controlling PSP amplitudes, is sufficiently small. For $\gamma b_1 > 4$, on the other hand, the mean dynamics will never converge.

\begin{figure}
\psfrag{e}{\raisebox{1mm}{\scriptsize $a_1$}}
\psfrag{gre}{\scriptsize $\quad\gamma b_1$}
\psfrag{g = 0.2}{\scriptsize $\!\!\!\!\gamma = 0.2$}
\psfrag{g = 3.7}{\scriptsize $\!\!\!\!\gamma = 3.7$}
\psfrag{a}{\raisebox{2mm}{\scriptsize $\mean{x}_{\infty}/\mean{s}$}}
\centerline{\includegraphics[width=0.7\textwidth]{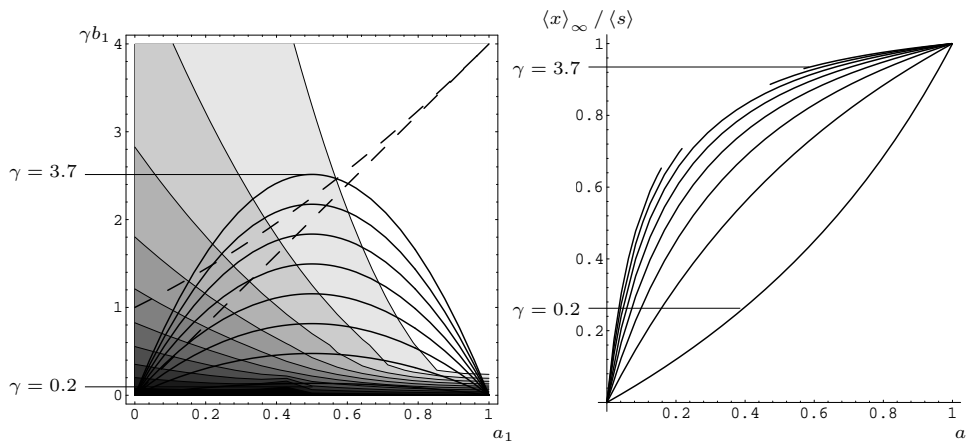}}
\caption[]{Left: Contour plot of the asymptotic mean membrane potential $\smean{x}_{\infty}$. Dashed lines delimit the regions of different mean dynamics shown in Fig.\ \ref{a1_b1_space}. Assuming Poisson statistics for stimulus times, the stimulus parameters $a_1$ and $b_1$ lie on parabolas, here plotted for $\gamma = 0.2,0.7,\ldots,3.7$. Right: Plot of $\smean{x}_{\infty}/\smean{s}$ for Poissonian stimulus times and the same values of $\gamma$ as on the left. The curves are interrupted where $(a_1 + 1)^2 < \gamma b_1$ such that the mean dynamics is divergent; cf.\ Fig.\ \ref{a1_b1_space}.
\label{asymp_<x>}}
\end{figure}

From Eq.\ (\ref{1.mom_iter}) we obtain the asymptotic values
\begin{eqnarray}
\label{<x>_infty}
\lim_{t \rightarrow \infty} \mean{V(t)} = \mean{x}_{\infty} & = & \frac{\gamma b_1}{\gamma b_1 + (1 - a_1)^2} \, \mean{s} \ , \\
\label{<y>_infty}
\mean{y}_{\infty} & = & \frac{\gamma (1 - a_1) a_1}{\gamma b_1 + (1 - a_1)^2} \, \mean{s}
\end{eqnarray}
for the regime of convergence. A contour plot of $\smean{x}_{\infty}$ as a function of $a_1$ and $\gamma b_1$ is shown in the left graph of Fig.\ \ref{asymp_<x>}. For the times $t_j$ of synaptic input being consistent with a Poisson process, it is shown in the Appendix that the stimulus parameters $a_1$, $b_1$ lie on a parabola, plotted in the left graph of Fig.\ \ref{asymp_<x>} for different $\gamma$. The ratio $\smean{x}_{\infty}/\smean{s}$ behaves then as shown in the right graph of the figure. Not surprisingly, the mean membrane potential is pulled closer to the mean synaptic reversal potential $\smean{s}$ with increasing $a_1$, that is, with increasing stimulus frequency, and with increasing PSP amplitude $\gamma$.

For $\gamma b_1 < 4 a_1$, the eigenvalues $\lambda_{1/2}$ are complex conjugate. In Sec.\ \ref{var_V}, I will show that only then will the variance of $V(t)$ converge. As depicted in Fig.\ \ref{a1_b1_space}, in this regime $\smean{V(t)}$ converges in a damped oscillation. The dynamics is solved straightforwardly. Let $L$ be the matrix that diagonalizes $M_1$, i.e., $L M_1 L^{-1}$ is diagonal. Furthermore, let
\begin{equation}
K := \left(\begin{array}{cc} 1 & 1 \\ {\rm i} & -{\rm i} \end{array}\right) \ .
\end{equation}
Then the matrix $KL$ is real and we can write the powers of $M_1$ as
\begin{equation}
\label{M_1^j(phi)}
M_1^j = a_1^j L^{-1} K^{-1} \left(\begin{array}{cc} \cos(j \phi) & -\sin(j \phi) \\ \sin(j \phi) & \cos(j \phi)
                                  \end{array}\right) KL \ , \quad\mbox{with}\quad \phi := {\rm arg}(\lambda_1) \ .
\end{equation}
The iteration dynamics (\ref{1.mom_iter}) is solved by
\begin{equation}
\label{spiral_motion}
{\mean{x}_j \choose \mean{y}_j} = {\mean{x}_{\infty} \choose \mean{y}_{\infty}}
- M_1^{j-1} \, {\mean{x}_{\infty} \choose \mean{y}_{\infty}} \ ,
\end{equation}
that is, a spiral motion around the attractive focus $(\mean{x}_{\infty},\mean{y}_{\infty})$. Its angular period is, measured in the number of synaptic input events,
\begin{equation}
\label{mean_osc_P}
P = \frac{2 \pi}{{\rm arg}(\lambda_1)} \ ,
\end{equation}
and averages in real time to
\begin{equation}
\label{mean_osc_T}
\mean{T} = P \mean{r} \ .
\end{equation}
Thus $\smean{T}$ is the mean period of $\smean{V(t)}$. Moreover, it may be shown easily that $P$ is the period of the covariance function,
\begin{multline}
\label{cov(xjxj+k)}
{\rm cov}(x_j,x_{j+k}) := \mean{\left(x_j - \mean{x}_j\right) \left(x_{j+k} - \mean{x}_{j+k}\right)} \\
= \int_{-\infty}^{\infty} dx \int_{-\infty}^{\infty} dy \int_{-\infty}^{\infty} dx' \int_{-\infty}^{\infty} dy' \,
p_k(x,y|x',y') \, p_j(x',y') \, \left(x - \mean{x}_{j+k}\right) \left(x' - \mean{x}_j\right) \ ,
\end{multline}
where
\begin{eqnarray}
p_1(x,y|x',y') & := & \int_{-\infty}^{\infty} d\bar{x} \int_{-\infty}^{\infty} d\bar{y} \,
p_{\cal R}(x,y|\bar{x},\bar{y}) \, p_{\cal S}(\bar{x},\bar{y}|x',y') \ ,\\
p_k(x,y|x',y') & := & \int_{-\infty}^{\infty} d\bar{x} \int_{-\infty}^{\infty} d\bar{y} \,
p_1(x,y|\bar{x},\bar{y}) \, p_{k-1}(\bar{x},\bar{y}|x',y') \quad
\mbox{for $k > 1$.} \nonumber
\end{eqnarray}
In particular, the asymptotic covariance function $\lim_{j \rightarrow \infty} {\rm cov}(x_j,x_{j+k})$ alternates between phases of correlation and anticorrelation with period $P$. In Sec.\ \ref{stat_fluct_osc_V}, I will show that under certain stimulus conditions these oscillations of the membrane potential never die out for {\em individual realizations} of the stochastic process. The damping of the {\em mean} oscillation is then due to a loss of phase coherence with time.

\begin{figure}
\psfrag{a1}{\scriptsize $a_1$}
\psfrag{r}{\scriptsize $\mean{r/\tau}$}
\psfrag{g=0.2}{\scriptsize $\gamma = 0.2$}
\psfrag{g=3.7}{\scriptsize $\gamma = 3.7$}
\psfrag{T}{\scriptsize $\mean{T/\tau}$}
\centerline{\includegraphics[width=0.7\textwidth]{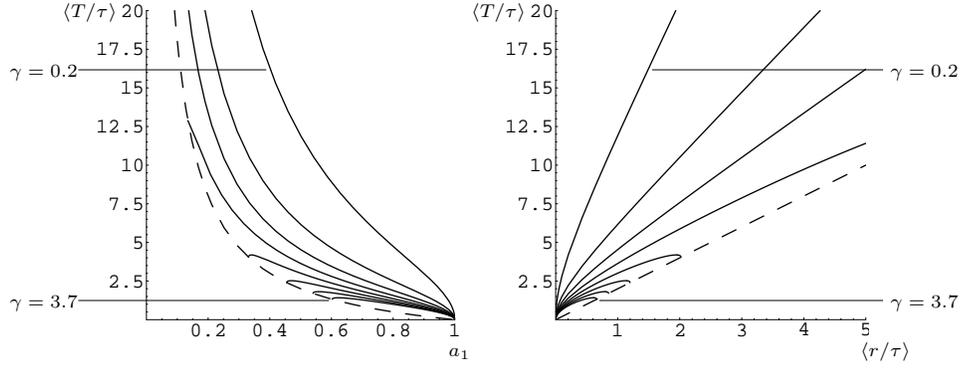}}
\caption[]{Mean oscillation period $\smean{T}$ of the membrane potential in units of the rise time $\tau$ of PSPs [cf.\ Eq.\ (\ref{PSP})], plotted as a function of the stimulus parameters $a_1$ (left) and $\smean{r/\tau}$ (right). For the curves we assume Poisson statistics for stimulus times and $\gamma = 0.2,0.7,\ldots,3.7$. The mean oscillation period lies on the dashed curves for $a_1 < 1 - 4/(e \gamma)$ or, equivalently, $\smean{r/\tau} > 4/(e \gamma - 4)$.
\label{osc_period}}
\end{figure}

For Poissonian stimulus times $t_j$, the mean oscillation period is given by
\begin{equation}
\label{mean_osc_T_Poisson}
\mean{\frac{T}{\tau}} =
\left\{
\begin{array}{ll}
2 \pi \mean{\frac{r}{\tau}} \Big/ \arctan\left[
\frac{\sqrt{e \gamma \mean{r/\tau} [(4 - e \gamma) \mean{r/\tau} + 4]}}{(2 - e \gamma) \mean{r/\tau} + 2} \right]
& \mbox{for $(2 - e \gamma) \mean{\frac{r}{\tau}} + 2 > 0$,} \\
2 \pi \mean{\frac{r}{\tau}} \Big/ \left\{ \pi + \arctan\left[
\frac{\sqrt{e \gamma \mean{r/\tau} [(4 - e \gamma) \mean{r/\tau} + 4]}}{(2 - e \gamma) \mean{r/\tau} + 2} \right] \right\}
& \mbox{elsewhere;}
\end{array}
\right.
\end{equation}
cf.\ the Appendix. Figure \ref{osc_period} shows plots of $\smean{T/\tau}$ for different $\gamma$, both as a function of $\smean{r/\tau}$ and $a_1$. For $\smean{r/\tau} > 4/(e \gamma - 4)$ or, equivalently, $a_1 < 1 - 4/(e\gamma)$, the stimulus enters the regime where $\lambda_{1/2}$ are real and negative, and the mean period ends up on the curve
\begin{equation}
\mean{\frac{T}{\tau}} = 2\mean{\frac{r}{\tau}} = 2 \left( \frac{1}{a_1} - 1 \right) \ ,
\end{equation}
plotted with the dashed lines in Fig.\ \ref{osc_period}. For $\smean{r/\tau} \rightarrow 0$ or, equivalently, $a_1 \rightarrow 1$, we find that $\smean{T/\tau}$ approaches zero. In particular, $\smean{T}$ can be much shorter than the rise time $\tau$ of PSPs.

\subsection{Variance of the membrane potential\label{var_V}}

\begin{figure}
\begin{minipage}{0.6\textwidth}
\leftline{\includegraphics[width=0.9\textwidth]{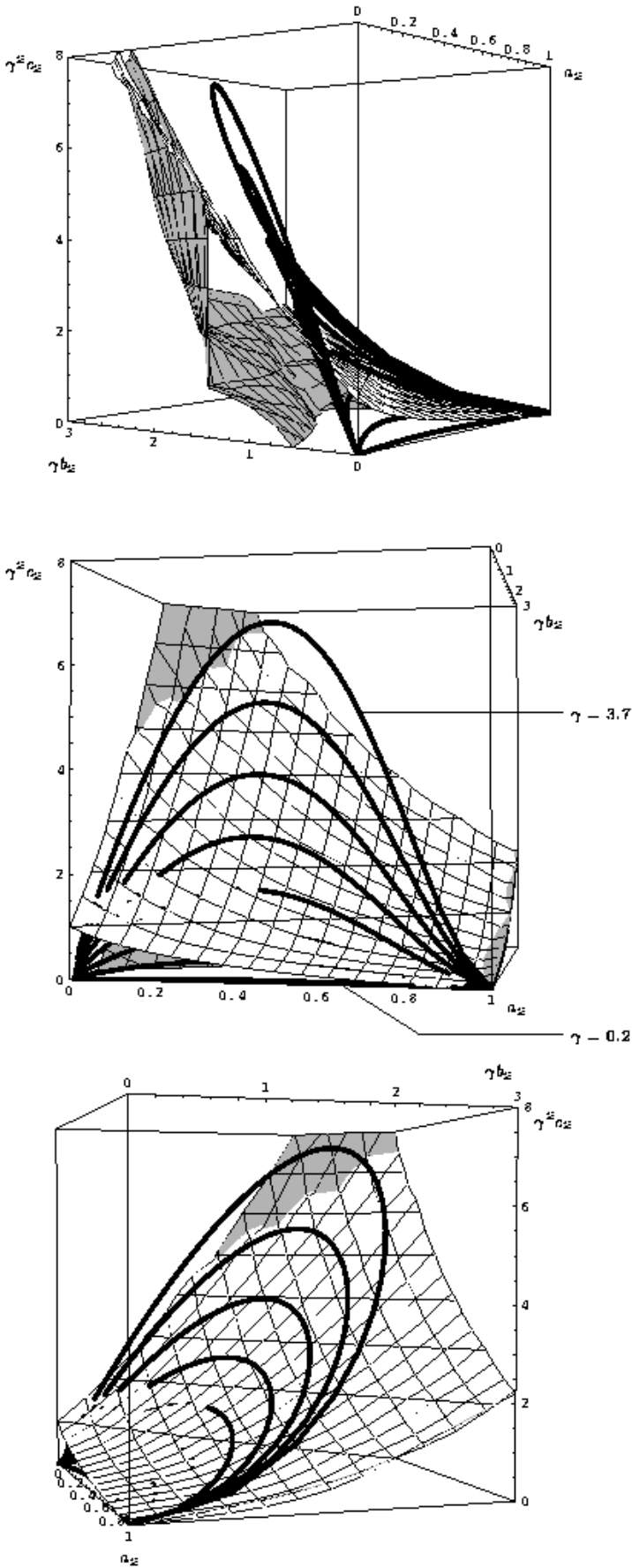}}
\end{minipage}
\begin{minipage}{0.39\textwidth}
\caption[]{Three different views of the two surfaces defined in Eq.\ (\ref{M2_constrain_surfaces_def}) in the space of stimulus parameters $a_2$, $b_2$, and $c_2$. The region of parameters that result in convergence of the second moment of the membrane potential is the space between the two surfaces that includes the $a_2$ axis. Parameters for Poissonian stimulus times lie on the thick curves for $\gamma = 0.2,0.7,\ldots,3.7$. The graphs show that convergence is ensured for all Poissonian stimuli, if $\gamma \le 1.2$.
\label{M2_constrain_surfaces_Poisson}}
\end{minipage}
\end{figure}

To estimate whether the trajectories $V(t)$ stay bounded when their mean converges to a finite value, we have to check whether their variance converges as well. We will now analyze the dynamic map for the second moments of $x$ and $y$ defined in Sec.\ \ref{Markov_form}. From Eqs.\ (\ref{momS_iter}) and (\ref{momR_iter}), we obtain
\begin{equation}
\label{2.mom_iter}
\left(\begin{array}{c} \mean{x^2}_j \\ \mean{xy}_j \\ \mean{y^2}_j \end{array}\right) =
\underbrace{\left(\begin{array}{ccc}
a_2 - \gamma b_2 + \gamma^2 c_2 & b_2 - 2 \gamma c_2 & c_2 \\
-\gamma a_2 + \frac{\gamma^2}{2} b_2 & a_2 - \gamma b_2 & \frac{1}{2} b_2 \\
\gamma^2 a_2 & -2 \gamma a_2 & a_2
\end{array}\right)}_{=: M_2}
\left(\begin{array}{c} \mean{x^2}_{j-1} \\ \mean{xy}_{j-1} \\ \mean{y^2}_{j-1} \end{array}\right)
+ \left(\begin{array}{c} u_{j-1} \\ v_{j-1} \\ w_{j-1} \end{array}\right) \ ,
\end{equation}
\[
\mean{x^2}_1 = \mean{x y}_1 = \mean{y^2}_1 = 0 \ ,
\]
with the stimulus parameters
\begin{equation}
\label{stim_params_2}
\left.\begin{array}{rcl}
a_2 & := & \mean{e^{-2r/\tau}} \\
b_2 & := & \mean{\frac{2 r}{\tau} e^{1 - 2r/\tau}} \\
c_2 & := & \mean{\left(\frac{r}{\tau}\right)^2 e^{2 - 2r/\tau}}
\end{array}\right\} \in (0,1) \ ,
\end{equation}
and
\begin{eqnarray}
u_j & := & \left(\gamma  b_2 - 2 {{\gamma }^2} c_2\right) \mean{s} \mean{x}_j + 2 \gamma  c_2 \mean{s} \mean{y}_j + {{\gamma }^2} c_2 \mean{s^2} \ ,\\
v_j & := & \left(\gamma a_2 - {{\gamma }^2} b_2\right) \mean{s} \mean{x}_j + \gamma  b_2 \mean{s} \mean{y}_j + \frac{1}{2}{{{\gamma }^2} b_2 \mean{s^2}} \ ,\\
w_j & := & - 2 {{\gamma }^2} a_2 \mean{s} \mean{x}_j + 2 \gamma a_2 \mean{s} \mean{y}_j + {{\gamma }^2} a_2 \mean{s^2} \ .
\end{eqnarray}
The $\smean{x}_j$ and $\smean{y}_j$ converge to the values given in Eqs.\ (\ref{<x>_infty}) and (\ref{<y>_infty}) such that $(u_j,v_j,w_j)$ will become constant. To check convergence of the second moments, it is thus necessary and sufficient to consider the eigenvalues of $M_2$. These are the roots of the characteristic polynomial
\begin{equation}
\label{M_2_charpol}
\nu^3 - \left(3 a_2 - 2 \gamma b_2 + \gamma^2 c_2\right) \nu^2
+ \left(3 a_2^2 - \gamma^2 a_2 c_2 - 2 \gamma a_2 b_2 + \frac{1}{2} \gamma^2 b_2^2\right) \nu - a_2^3 = 0 \ ,
\end{equation}
and are rather lengthy expressions which need not be spelled out here. Depending on the stimulus parameters $a_2$, $b_2$, and $c_2$, there are one real and two complex conjugate eigenvalues, or three real eigenvalues. Let $\nu_1$ be the eigenvalue that is always real and $\nu_{2/3}$ the other two that may be complex conjugate or real. Stimulus parameters $a_2$, $b_2$, $c_2$ that yield a convergent second moment of $V(t)$ are those that obey the constraints
\begin{equation}
\left| \nu_1 \right| =: f_1(a_2,\gamma b_2,\gamma^2 c_2) < 1 \ ,\quad
\max \left( \left| \nu_2 \right| , \left| \nu_3 \right| \right) =: f_2(a_2,\gamma b_2,\gamma^2 c_2) < 1 \ ,
\end{equation}
with continuous functions $f_1$ and $f_2$. The two surfaces defined
by
\begin{equation}
\label{M2_constrain_surfaces_def}
f_1(a_2,\gamma b_2,\gamma^2 c_2) = 1 \ ,\quad
f_2(a_2,\gamma b_2,\gamma^2 c_2) = 1
\end{equation}
are shown in Fig.\ \ref{M2_constrain_surfaces_Poisson}. Since convergence is obviously ensured for $\gamma = 0$, which yields $x_j \equiv y_j \equiv 0$ [cf.\ Eq.\ (\ref{map_dyn})], the parameter region that results in convergence of the second moments is the space {\em between} the two surfaces that includes the axis $(a_2,\gamma b_2,\gamma^2 c_2)  = (a_2,0,0)$, $a_2 \in (0,1)$. In the region beyond the intersection of the surfaces, i.e., for roughly $\gamma^2 c_2 > 9$, there are no combinations of parameters that yield convergent second moments.

For Poisson statistics of the stimulus times $t_j$, the stimulus parameters $a_2$, $b_2$, $c_2$ lie on the curves plotted in Fig.\ \ref{M2_constrain_surfaces_Poisson} for different values of $\gamma$; cf.\ the Appendix. The curves run from $(a_2,\gamma b_2,\gamma^2 c_2) = (0,0,0)$, the limiting point for low input activity ($\smean{r} = \infty$), to $(a_2,\gamma b_2,\gamma^2 c_2) = (1,0,0)$, the limit of high-frequency stimulation ($\smean{r} = 0$). For $\gamma$ sufficiently small, the curves lie completely within the region of convergence. For larger $\gamma$, they are in the region of divergence except near the point $(a_2,\gamma b_2,\gamma^2 c_2) = (0,0,0)$. For $\smean{r/\tau} \ll 1$, which is the realistic regime for cortical neurons, the eigenvalues $\nu_{2/3}$ are complex conjugate and we get
\begin{eqnarray}
\label{M_2_eigenvalues_1}
\nu_1 & = & 1 - \left( 2 - \frac{e \gamma}{2} \right) \mean{\frac{r}{\tau}} + O\!\left(\mean{\frac{r}{\tau}}^{3/2}\right) \ ,\\
\label{M_2_eigenvalues_2/3}
\left|\nu_{2/3}\right| & = & 1 - \left( 2 + \frac{e \gamma}{4} \right) \mean{\frac{r}{\tau}} + O\!\left(\mean{\frac{r}{\tau}}^{3/2}\right) \ .
\end{eqnarray}
It follows that for $\smean{r/\tau} \ll 1$, it is necessary and sufficient for the second moments to converge that $\gamma < 4/e$. In fact, Fig.\ \ref{M2_constrain_surfaces_Poisson} shows that at least for $\gamma \le 1.2$ the second moments converge for all $0 < \smean{r/\tau} < \infty$, corresponding to the entire curves running between $(a_2,\gamma b_2,\gamma^2 c_2)  = (0,0,0)$ and $(a_2,\gamma b_2,\gamma^2 c_2) = (1,0,0)$ in the parameter space.

As shown in the Appendix, the condition $\gamma < 4/e$ is for Poisson statistics of the times $t_j$ equivalent to $4 a_1 > \gamma b_1$ for all $a_1 \in (0,1)$. In the following, we will assume this condition to hold. The system is thus always in the regime of damped oscillations of $\smean{V(t)}$; cf.\ Fig.\ \ref{a1_b1_space}.

After some lengthy but straightforward algebra, we find for the asymptotic variance of $x$, and hence of $V(t)$,
\begin{equation}
\label{var(x)_infty}
\lim_{t \rightarrow \infty} {\rm var}\!\left[V(t)\right] = {\rm var}_{\infty}(x) := \mean{x^2}_{\infty} - \mean{x}_{\infty}^2 =
\mean{s^2} \rho_1 - \mean{s}^2 \rho_2 \ ,
\end{equation}
with coefficients
\begin{eqnarray}
\rho_1 & = & \frac{{{\gamma }^2} \left( {{b_2}^2} + 2 c_2 - 
       2 a_2 c_2 \right) }{2 {{\left( 1 - a_2 \right) }^3} + 
     4 \gamma \left( 1 - a_2 \right) b_2 + 
     {{\gamma }^2} {{b_2}^2} - 
     2 {{\gamma }^2} \left( 1 + a_2 \right) c_2} \ ,\\
\rho_2 & = & \frac{{{\gamma }^2} {{b_1}^2}}{{{\left[ {{\left( 1 - a_1 \right) }^2} + \gamma  b_1 \right] }^2}}\\
&& - \, \frac{2 {{\gamma }^2} \left[ a_1 \left( 1 - a_1 \right)
         \left( {{b_2}^2} + 2 c_2 - 2 a_2 c_2
            \right)  + b_1 
         \left( b_2 - a_2 b_2 - 2 \gamma  c_2
            \right)  \right] }{\left[ {{\left( 1 - a_1 \right) }^2} + \gamma  b_1 \right]  
      \left[ 2 {{\left( 1 - a_2 \right) }^3} + 
        4 \gamma \left( 1 - a_2 \right) b_2 + 
        {{\gamma }^2} {{b_2}^2} - 
        2 {{\gamma }^2} \left( 1 + a_2 \right) c_2 \right] } \ .\nonumber
\end{eqnarray}
For Poisson statistics of the stimulus times $t_j$, the coefficients $\rho_{1/2}$ simplify to
\begin{eqnarray}
\rho_1 & = & \frac{(e \gamma)^2}{4 e \gamma - (e \gamma)^2 + 4 \mean{r/\tau}}
> 0 \ , \\
\rho_2 & = & \frac{(e \gamma)^3 \left(e \gamma + 2 \mean{r/\tau}\right)}{\left[4 e \gamma
- (e \gamma)^2 + 4 \mean{r/\tau}\right] \left(e \gamma + \mean{r/\tau}\right)^2}
> 0 \ ;
\end{eqnarray}
cf.\ the Appendix.

\subsection{Stationary states, fluctuations, and noise-driven oscillations\label{stat_fluct_osc_V}}

We have seen in the two previous sections that there is a region of stimulus parameters where the mean and variance of $V(t)$ converge to finite values. Averages do not tell us, however, what {\em individual} trajectories $V(t)$ look like. In this section we want to gain insight into the temporal pattern of individual trajectories.

\begin{figure}
\psfrag{t}{\scriptsize $t/\tau$}
\psfrag{V}{\scriptsize $V$}
\psfrag{0}{\scriptsize 0}
\psfrag{s}{\scriptsize $\mean{s}$}
\psfrag{ds = 0.1 s}{\scriptsize $\Delta s = 0.1 \, \mean{s}$}
\psfrag{ds = s}{\scriptsize $\Delta s = \mean{s}$}
\psfrag{ds = 10 s}{\scriptsize $\Delta s = 10 \, \mean{s}$}
\psfrag{r = 0.001}{\scriptsize $\mean{r/\tau} = 0.001$}
\psfrag{r = 0.01}{\scriptsize $\mean{r/\tau} = 0.01$}
\psfrag{r = 0.1}{\scriptsize $\mean{r/\tau} = 0.1$}
\centerline{\includegraphics[width=0.9\textwidth]{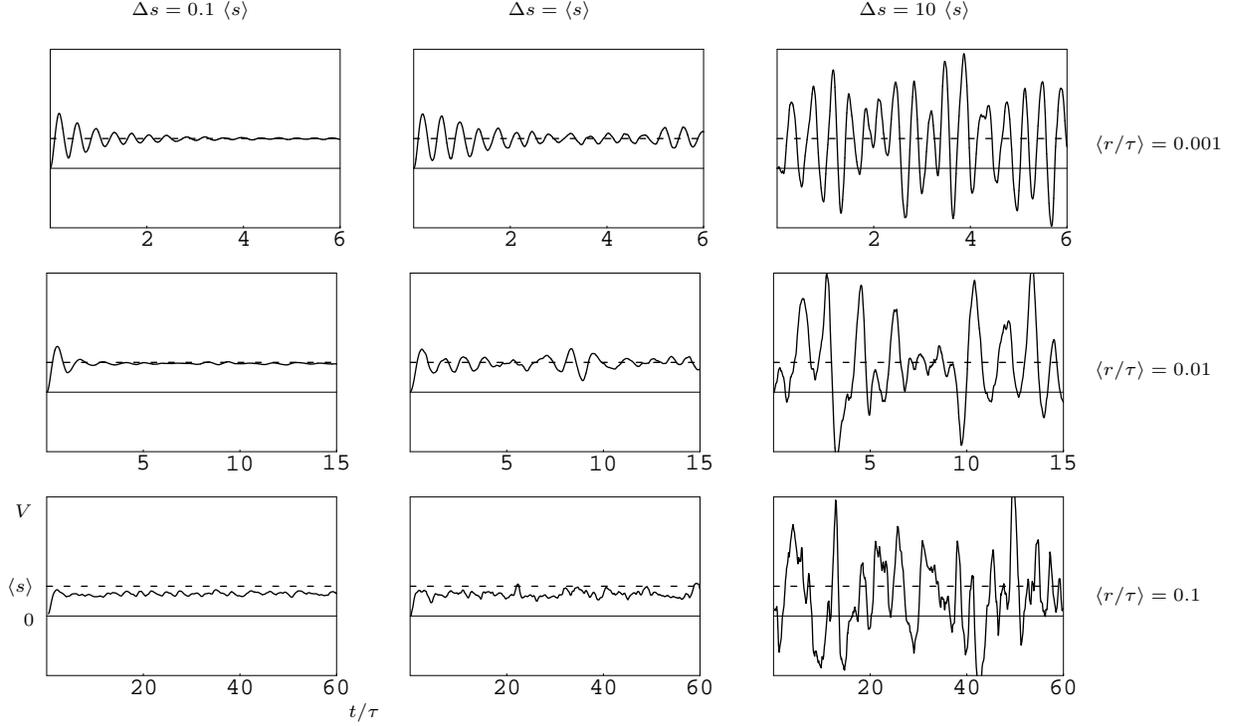}}
\caption[]{Typical trajectories of the membrane potential $V(t)$, simulated for various stimulus conditions as indicated by the row and column labels of the array of graphs. The unit of time is the rise time $\tau$ of PSPs [cf.\ Eq.\ (\ref{PSP})]. Synaptic reversal potentials are uniformly distributed in the intervals $[\smean{s} - \Delta s,\smean{s} + \Delta s]$. The membrane potentials $V = 0$ (resting potential) and $V = \smean{s}$ are indicated in each graph by the solid and dashed lines, respectively. The graphs show the transitions between stationary, fluctuating, and oscillatory dynamics of $V(t)$ as discussed in the main text. The PSP-amplitude factor $\gamma = 0.1$ for all graphs.
\label{stat_fluct_osc_transit}}
\end{figure}

Let us first deal with the short-time behavior of individual trajectories $(x_j,y_j)$. We ask what they look like for the first few $j$, that is, the first few synaptic inputs. The variances
\begin{equation}
{\rm var}_j(x) := \mean{x^2}_j - \mean{x}_j^2 \ , \quad {\rm var}_j(y) := \mean{y^2}_j - \mean{y}_j^2
\end{equation}
are zero initially. They increase to finite values no faster than the fastest-growing linear combination of second moments, i.e., like $e^{-j/Q}$ with
\begin{equation}
Q = -1 \left/ \ln\left(\displaystyle\min_{i = 1,2,3} \left|\nu_i\right|\right)
\right. \ .
\end{equation}
We have to compare $Q$ to the period $P$ of the oscillation of the mean values $(\smean{x}_j,\smean{y}_j)$ in order to see whether this oscillation shows up in individual realizations $(x_j,y_j)$. From Eqs.\ (\ref{mean_osc_P}), (\ref{M_2_eigenvalues_1}), and (\ref{M_2_eigenvalues_2/3}) we obtain
\begin{equation}
\frac{P}{Q} = \frac{(8 + e \gamma) \pi}{2 (e \gamma)^{1/2}} \mean{\frac{r}{\tau}}^{1/2}
 + O\!\left(\mean{\frac{r}{\tau}}\right) \ .
\end{equation}
Thus for $\smean{r/\tau}$ sufficiently small, we get $P/Q \ll 1$ and the oscillation of the means $\smean{x}_j$, $\smean{y}_j$ is fast as compared to the growth time of the fluctuations ${\rm var}_j(x)$, ${\rm var}_j(y)$ around the means. Individual realizations $(x_j,y_j)$ are then well described by their means for several periods of the oscillation. Put differently, an oscillation with a mean period given by Eq.\ (\ref{mean_osc_T_Poisson}) then shows up in individual realizations $V(t)$. With longer interstimulus times $\smean{r/\tau}$, fluctuations increasingly interfere with the oscillation. The transition from an oscillation-dominated to a fluctuation-dominated dynamics of $V(t)$ is depicted in Fig.\ \ref{stat_fluct_osc_transit}.

It remains to establish the long-time behavior of trajectories $(x_j,y_j)$. The dynamics (\ref{spiral_motion}) of their mean value spirals into the point $(\smean{x}_{\infty},\smean{y}_{\infty})$. Without damping of the oscillation, the trajectories would lie on orbits defined by $q(x - \smean{x}_{\infty}, y - \smean{y}_{\infty}) = {\rm const}$, with the quadratic form
\begin{equation}
q(\xi, \eta) := \skap{(\xi, \eta)}{(KL)^{\dag} KL (\xi, \eta)}
= \frac{4 \gamma a_1^2 \xi^2}{b_1}
- 4 \gamma a_1 \xi \eta
+  4 a_1 \eta^2 \ .
\end{equation}
To estimate the true degree of damping of individual trajectories $(x_j, y_j)$, we calculate the mean asymptotic ratio $\smean{q/q_0}_{\infty}$ with the initial value $q_0 := q(\smean{x}_{\infty},\smean{y}_{\infty})$ of the quadratic form $q$. From Eq.\ (\ref{2.mom_iter}) we obtain the three second moments $\smean{x^2}_{\infty}$, $\smean{xy}_{\infty}$, and $\smean{y^2}_{\infty}$ which are needed for the calculation of $\smean{q}_{\infty}$. After some lengthy but straightforward algebra, we find
\begin{equation}
\mean{\frac{q}{q_0}}_{\infty} =  \frac{\mean{s^2}}{\mean{s}^2} \bar{\rho}_1
- \bar{\rho}_2 \ ,
\end{equation}
where the coefficients are for Poisson statistics of synaptic input times $t_j$,
\begin{eqnarray}
\bar{\rho}_1 & = & \frac{2 \left(e \gamma + \mean{r/\tau}\right)^2}{4 e \gamma - (e \gamma)^2 + 4 \mean{r/\tau}}
> 0 \ ,\\
\bar{\rho}_2 & = & \frac{2 e \gamma (e \gamma + 2 \mean{r/\tau})}{4 e \gamma - (e \gamma)^2 + 4 \mean{r/\tau}}
> 0 \ ;
\end{eqnarray}
cf.\ the Appendix. For ${\rm var}(s) = \smean{s^2} - \smean{s}^2 = 0$, it follows that
\begin{equation}
\mean{\frac{q}{q_0}}_{\infty} = \bar{\rho}_1 - \bar{\rho}_2
= \frac{2 \mean{r/\tau}^2}{4 e \gamma - (e \gamma)^2 + 4 \mean{r/\tau}}
< \frac{1}{2} \mean{\frac{r}{\tau}} \ll 1 \ .
\end{equation}
Hence there is strong damping, and individual trajectories $(x_j,y_j)$ converge close to the steady mean state, if synaptic currents have all the same reversal potential. On the other hand, for ${\rm var}(s)/\smean{s}^2 \gg 1$, hence $\smean{s^2}/\smean{s}^2 \gg 1$, we get $\smean{q/q_0}_{\infty} \gg 1$ and there is no damping of individual trajectories $(x_j,y_j)$. Since the dynamics is a temporally homogeneous Markov chain, at any time we then find qualitatively the same situation as at the start of the process. Thus there is no qualitative change in the trajectories $(x_j,y_j)$ on a long time scale, and the pattern of evolution, random fluctuations or oscillations, that dominates initially (see above) will also prevail at all times. Figure \ref{stat_fluct_osc_transit} summarizes the types of dynamics of $V(t)$, illustrating our results on short- and long-time behavior by computer simulations.

\begin{figure}
\psfrag{r}{\scriptsize $\mean{r/\tau}$}
\psfrag{T}{\scriptsize $\mean{T/\tau}$}
\psfrag{D}{\scriptsize $\mean{\Delta/\tau}$}
\centerline{\includegraphics[width=0.5\textwidth]{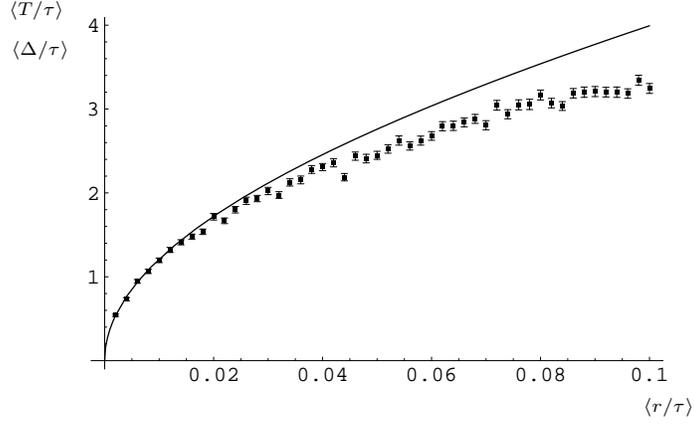}}
\caption[]{Comparison of the mean oscillation period $\smean{T}$ of the membrane potential as given by Eq.\ (\ref{mean_osc_T_Poisson}) (solid line; cf.\ Fig.\ \ref{osc_period}) with the mean jitter period $\smean{\Delta}$ defined in Eq.\ (\ref{def_jitter_period}) as observed in computer simulations (box symbols; bars indicate standard errors). The unit of time is the rise time $\tau$ of PSPs [cf.\ Eq.\ (\ref{PSP})]. The match between the two periods is perfect for small $\smean{r/\tau}$, when oscillations are rather regular. As oscillations are increasingly degraded by fluctuations for larger $\smean{r/\tau}$ (cf.\ Fig.\ \ref{stat_fluct_osc_transit}), the mean jitter period drops below the mean oscillation period. In the simulations, synaptic reversal potentials are uniformly distributed in an interval with $\smean{s} = 0$; the PSP-amplitude factor $\gamma = 0.1$.
\label{osc_jitter_period_comp}}
\end{figure}

With synaptic reversal potentials $s_j$ having a high variance, we have seen individual trajectories $V(t)$ to oscillate or fluctuate persistently around the value $\lim_{t \rightarrow \infty} \smean{V(t)} = \smean{x}_{\infty}$. It is interesting to compare the mean of the intervals $\Delta = t - t'$ between successive times $t > t'$ defined by
\begin{equation}
\label{def_jitter_period}
V(t) = V(t') = \mean{x}_{\infty} \ ,\quad \frac{d}{dt} V(t) > 0 \ ,\quad \frac{d}{dt} V(t') > 0 \ ,
\end{equation}
the mean ``jitter period'', with the mean oscillation period $\smean{T}$ [cf.\ Eq.\ (\ref{mean_osc_T_Poisson})] of $\smean{V(t)}$. I have measured jitter periods in computer simulations of $V(t)$. As can be seen in Fig.\ \ref{osc_jitter_period_comp}, the match between the two periods is perfect for small $\smean{r/\tau}$, that is, in the regime where oscillations are rather regular. For increasing interstimulus times $\smean{r/\tau}$, when the random-walk component of membrane dynamics grows stronger (cf.\ Fig.\ \ref{stat_fluct_osc_transit}), the mean jitter period drops below the mean oscillation period, indicating that fluctuations cause $V(t)$ to jitter around its asymptotic mean value faster than the oscillatory component of the dynamics alone.

\subsection{Delays\label{delays}}

The conduction of synaptic currents in neuronal dendrites leads to delays relative to the time of the synaptic input. Let us assume here that we can assign a delay $d_i > 0$ to a PSP initiated at time $t_i$, such that at time $t_i + d_i$ the response is spread out across the whole neuron. Of course, such a delay does not properly describe gradual PSP propagation. In a sense, it is the opposite extreme of the instantaneous PSP propagation that we have considered so far. The dynamics of the membrane potential with such delayed PSPs is given by
\begin{equation}
\label{V_tot_delayed}
V(t) = \sum_{i=1}^{\infty} \Lambda\!\left(\gamma_i,s_i,t_i;t - d_i\right) \ ;
\end{equation}
cf.\ Eq.\ (\ref{V_tot}). A reformulation as a Markov chain as in Sec.\ \ref{Markov_form} is now not possible. This fact calls for a reconsideration of our previous results. Here I am concerned with proving structural stability of the dynamics analyzed above with respect to small delay perturbations. To this end, we may extend the previous dynamics to incorporate first-order delay effects. The issue of delays is covered in detail in \cite{Hillenbrand01} for a slightly more general class of dynamical system. In this paper, I only sketch the way to proceed.

Expanding Eq.\ (\ref{V_tot_delayed}) to first order in the delays $d_i/\tau$, we have to note that $\Lambda(\gamma_i,s_i,t_i;t)$ is not differentiable at $t = t_i$; cf.\ Eq.\ (\ref{PSP}). We can take advantage of the fact, however, that $d_i > 0$ and write
\begin{eqnarray}
\label{PSP_expansion}
\lefteqn{\Lambda(\gamma_i,s_i,t_i;t - d_i)} && \\
& = & \Lambda(\gamma_i,s_i,t_i;t) + d_i \, \lim_{d \rightarrow 0+}
\frac{\Lambda(\gamma_i,s_i,t_i;t) - \Lambda(\gamma_i,s_i,t_i;t - d)}{d} + O(d_i^2) \nonumber\\
& = & \Lambda(\gamma_i,s_i,t_i;t) + \frac{d_i}{\tau} \,
\left( 1 - \frac{t}{\tau} \right) e^{1-t/\tau} \, \Theta(t - t_i)
+ O\!\left[\left(\frac{d_i}{\tau}\right)^2\right] \ ,\nonumber
\end{eqnarray}
that is, we take the derivative of $\Lambda(\gamma_i,s_i,t_i;t)$ from lower values of $t$. Equation (\ref{PSP_expansion}) is substituted into the dynamic Eq.\ (\ref{V_tot_delayed}) and only terms up to first order in $d_i/\tau$ are kept. As before, we use $\gamma \equiv \gamma_i$ to obtain a model with a minimal set of variables. We can now transform to new dynamic variables $x_j := V(t_j)$, $y_j$, and $z_j$ that obey the stochastic iteration
\begin{equation}
\label{map_dyn_delay}
\left(\begin{array}{c} x_j \\ y_j \\ z_j \end{array}\right) = {\cal R}'(r_{j - 1}) \circ {\cal S}'(s_{j - 1},d_{j - 1}) \,
\left(\begin{array}{c} x_{j - 1} \\ y_{j - 1} \\ z_{j - 1} \end{array}\right) \ , \quad x_1 = y_1 = z_1 = 0 \ ,
\end{equation}
\begin{eqnarray}
\label{S_map_delay}
{\cal S}'(s,d): && \left(\begin{array}{c} x \\ y \\ z \end{array}\right) \mapsto
\left(\begin{array}{c}
x \\
y + \gamma (s - x) \left( 1 + \frac{d}{\tau} \right) \\
z + \gamma (s - x) \frac{d}{\tau}
\end{array}\right) \ ,\\
\label{R_map_delay}
{\cal R}'(r): && \left(\begin{array}{c} x \\ y \\ z \end{array}\right) \mapsto
\left(\begin{array}{c}
\left( x + e y \frac{r}{\tau} - e z \right) e^{-r/\tau} \\
y e^{-r/\tau} \\
z e^{-r/\tau}
\end{array}\right) \ .
\end{eqnarray}
The dimension of the stochastic dynamic map is increased by one compared to the case without delays; cf.\ Eq.\ (\ref{map_dyn}). Treating Eq.\ (\ref{map_dyn_delay}) analogous to Eq.\ (\ref{map_dyn}), we can derive dynamic maps for the moments of $x$, $y$, and $z$. These will have accordingly higher dimensions than those for the moments of $x$ and $y$ without delays. This underlines the necessity to check the structural stability of the dynamics derived previously.

It can be shown \cite{Hillenbrand01} that the dynamics for the first and second moments of $x$ and $y$ is stable with respect to small delay perturbations, provided that
\begin{equation}
\label{delay_cond}
\gamma b_2 < 2 (a_2 + 1)^2 \ .
\end{equation}
In general, this is a condition for convergence in the delayed system that is additional to those derived for the undelayed system. For Poisson statistics of synaptic input times, however, we know that $a_2$ and $b_2$ lie on the parabola $b_2 = e a_2 (1 - a_2)$; see the Appendix. Together with the condition $\gamma < 4/e$ derived in Sec.\ \ref{var_V} for the convergence of the variance of $V(t)$, this implies condition (\ref{delay_cond}).

By continuity of eigenvalues and asymptotic values in the delays within the extended model, it follows that for small delays there is only a small quantitative and no qualitative change in membrane dynamics. All that has been concluded on patterns of the dynamics hence remains true for small delays. Moreover, it can be shown that small delays decrease the asymptotic attraction of $\smean{V(t)}$ to the mean synaptic reversal potential $\smean{s}$ and increase the mean period $\smean{T/\tau}$ of membrane oscillations \cite{Hillenbrand01}.

\section{Summary and discussion}

In this paper, I have analyzed the subthreshold dynamics of the neural membrane potential driven by stochastic synaptic input of stationary statistics. Conditions on the input statistics for stability of the dynamics have been derived. Regimes of input statistics for stationary, fluctuating, and oscillatory dynamics have been identified. For the case of Poissonian stimulus times, that is, temporal noise, it has turned out that persistent oscillations can develop with a mean period that depends nontrivially on the mean interstimulus time. In particular, noise-driven oscillations occur in the absence of any pace-making mechanism in the stimulus, in the intrinsic neural dynamics, or in a recurrent neural network.

What does it mean for a real neuron, if its membrane potential is ``unstable'' under stimulation by the network's synaptic input? As the analysis has shown, instability of the first or second moments implies excursions of $V(t)$ with growing positive and negative amplitudes. After some stochastic period of time, therefore, the membrane potential will certainly cross the threshold for firing. The neuron will then be set to a post-spike potential that depends in some way on the stimulus history and the process will resume.

I have neglected many effects in the modeling for the sake of analytical feasibility. Most notably, PSPs have been given a shape that does not properly reflect conduction in neuronal dendrites. One shortcoming is a lack of variability of PSP shape; see, however, \cite{Chitwood_etal99,Jaffe&Carnevale99,Destexhe&Pare99}. Another is that real neuronal membranes contain ionic conductances which are voltage-gated \cite{Hille92}. Their effect is to modify the shape of PSPs in a voltage-dependent manner as they are propagated along a dendrite; see, e.g., \cite{Andreasen&Lambert99}. Moreover, with voltage-gated channels, PSPs do not simply add up but interact nonlinearly. The conclusions drawn in the present paper, therefore, can only be on qualitative system behavior and should not be understood quantitatively.

In the analyzed model, there is no representation of the spatial dimensions of a neuron. For a neuron where spatial conduction times are significant, the present results suggest that spatiotemporal waves of membrane potential develop in the regime of noise-driven oscillations. For the generation of action potentials, however, all that matters is the potential at the cell's soma.

\subsection{Oscillations in stochastic systems}

It is a common example in textbooks on stochastic dynamical systems to calculate stationary densities for a damped harmonic oscillator subject to an external stochastic force; see, e.g., \cite{Honerkamp94}. If the damped oscillator is in the periodic regime, the intrinsic oscillations are sustained by the stochastic force. In the context of biological systems, stochastically sustained oscillations have been analyzed, somewhat heuristically, for the population dynamics of an epidemic model \cite{Aparicio&Solari01}. This system is autonomous and an intrinsic oscillator. The stochastic nature of the dynamics prevents asymptotic convergence to a steady state.

It is thus a known generic property of periodic relaxation systems to exhibit oscillations at their intrinsic frequency, sustained by some stochastic influence. The dynamics analyzed in this paper, however, represents a different type of phenomenon. The system studied is not an intrinsic oscillator but exhibits oscillations at a mean period that is, up to a temporal scale, determined by the stochastic drive alone. The system can be formally viewed as a control loop where a sequence of brief signals (the synaptic reversal potentials $s_j$) controls via a slow response (the PSPs) a dynamic variable [the membrane potential $V(t)$]. The theme of the control loop is fully developed in \cite{Hillenbrand&vanHemmen00,Hillenbrand01}.

\subsection{Oscillations in neural systems}

Oscillations of membrane potential and spiking activity are quite ordinary in the neural systems of the brain. They arise under various conditions, with varying degree of correlation between neurons, and in a wide range of frequencies. Their functional implications may be equally various and are much debated today.

Explanations of oscillations have been basically of two kinds. One is in terms of intrinsic oscillator cells that act as a pacemaker for rhythmic activity in the network \cite{Steriade_etal93,Gray&McCormick96}. The other makes reference to the fact that recurrent neural networks have a natural tendency to produce rhythmic and synchronized activity \cite{Gerstner_etal96,Brunel&Hakim99}.

Some of neural oscillations are most probably generated by the intrinsic neural dynamics of ion channels. Others are propagated by synaptic potentials and are of less certain origin. A prominent example of the latter kind are cortical oscillations in the gamma frequency band (roughly 20--90 Hz). Cortical gamma oscillations are mostly evoked by a sensory stimulus. Thus, spontaneous activity in the visual cortex of awake cats and primates is rarely oscillatory, whereas visual stimuli of increasing speed of motion produce subthreshold and suprathreshold oscillations of increasing frequency \cite{Jagadeesh_etal92,Bringuier_etal97,Castelo-Branco_etal98,Gray&VianaDiPrisco97,Friedman-Hill_etal00,Maldonado_etal00}.

The results presented here suggest that oscillations of the neural membrane potential can arise from the network's background activity. Let us assume that a stimulus evokes responses in neurons of a coupled system at a rate that increases with stimulus speed, because more neurons in the network are stimulated per time at higher speeds \footnote{Different individual neurons respond best at different stimulus speeds, a property called speed tuning. For the complete coupled system, however, we may neglect effects of speed tuning.}. The observed dependence of oscillations on a stimulus is then predicted by Sec.\ \ref{stat_fluct_osc_V}, the relation between oscillation period and stimulus speed by Eq.\ (\ref{mean_osc_T_Poisson}). Note that the conditions $\smean{r/\tau} \ll 1$ and ${\rm var}(s)/\smean{s}^2 \gg 1$ for the development of noise-driven oscillations are probably fulfilled under external stimulation of a network of cortical neurons, each receiving roughly 10000 synapses of both an excitatory and inhibitory kind \cite{Defelipe&Farinas92}. The degree of correlation between neurons that is to be expected from noise-driven oscillations increases with the extent to which they share common input from the network's background activity. Correlations should, therefore, decrease with distance between neurons, in agreement with what is generally observed. In a network of spike-exchanging neurons, however, correlations can even arise between neurons that do not share any input.

The present analysis draws attention to a phenomenon, noise-driven oscillations, that should be very common in neural systems and may be the cause of some of the observed membrane-potential oscillations.

\subsection{Hyperpolarization-induced activity}

\begin{figure}
\psfrag{t}{\scriptsize $t/\tau$}
\psfrag{V}{\scriptsize $V$}
\psfrag{0}{\scriptsize 0}
\psfrag{sdep}{\scriptsize $s_{\rm dep}$}
\psfrag{shyp}{\scriptsize $s_{\rm hyp}$}
\centerline{\includegraphics[width=0.9\textwidth]{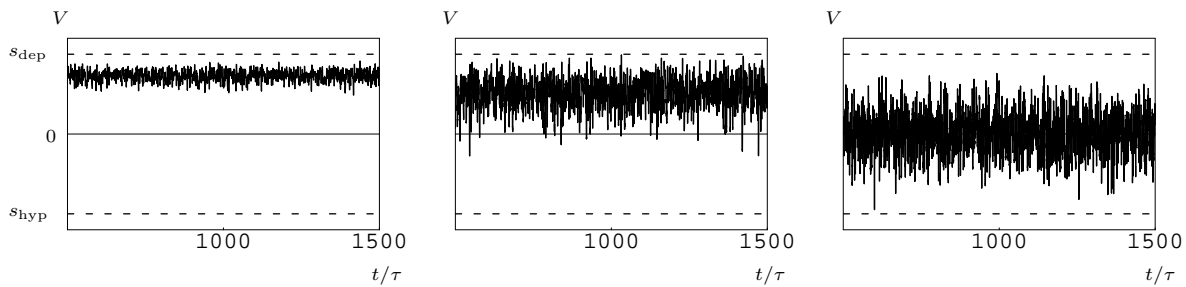}}
\caption[]{Demonstration of hyperpolarization-induced activity. The three graphs show long-time simulations of the membrane potential $V(t)$. Transients at the start of the simulation are cut off. The unit of time is the rise time $\tau$ of PSPs [cf.\ Eq.\ (\ref{PSP})]. Depolarizing synaptic input is applied with mean interstimulus time $\smean{r/\tau} = 0.1$ and reversal potential $s_{\rm dep} > 0$ as indicated by the upper dashed line in each graph. There is no hyperpolarizing synaptic input for the left graph; for the central graph there is hyperpolarization with $\smean{r/\tau} = 0.5$; for the right graph with $\smean{r/\tau} = 0.1$. The hyperpolarizing reversal potential is $s_{\rm hyp} = -s_{\rm dep} < 0$ as indicated by the lower dashed line in each graph. If the threshold for spike generation is assumed close to $s_{\rm dep}$, there will be no spikes for the case without hyperpolarizing input (left), a few spikes for weak hyperpolarizing input (center), and again no spikes for equal hyperpolarizing and depolarizing input (right). The PSP-amplitude factor $\gamma = 0.1$ for all graphs.
\label{hyperpol_exc}}
\end{figure}

There is an interesting consequence of the analytical results. It is, at first sight, somewhat counterintuitive. Consider a neuron that receives depolarizing synaptic input at a fixed average rate. Let us assume that at this level of depolarization the membrane potential remains mostly below the threshold for spike generation. Now, if we add some {\em hyper}polarizing synaptic input, it turns out that the neuron may actually {\em start spiking}. Further increase of the hyperpolarizing input rate eventually shuts neural activity off. This scenario is demonstrated in computer simulations shown in Fig.\ \ref{hyperpol_exc}.

The effect seems to be at odds with the usual notion of hyperpolarizing synapses to {\em inhibit} neural activity rather than {\em promote} it. Exceptions have only been reported for cases where a hyperpolarization-activated current repolarizes the cell, giving rise to a rebound burst of action potentials; see, e.g., \cite{Huguenard&McCormick92,McCormick&Huguenard92}. The effect demonstrated here is of a different nature. It results from an increase of membrane fluctuations with the addition of hyperpolarizing synaptic input; cf.\ Eq.\ (\ref{var(x)_infty}). For a range of hyperpolarizing input rates, increased fluctuations are likely to spontaneously overcome the associated drop in mean membrane potential; cf.\ Eq.\ (\ref{<x>_infty}). The result is fluctuation-driven spike generation.

The phenomenon of hyperpolarization-induced activity offers a subtle way in which neural spiking may be controlled. Whether it is actually used in the brain is unexplored today.

\appendix
\section{}

It is reasonable to assume the times $t_j$ at which synaptic inputs are received by a cortical neuron from other cortical neurons to obey Poisson statistics. For the density $u$ of interstimulus times $r$ this means
\begin{equation}
\label{exp_density}
u(r) = \frac{e^{-r/\mean{r}}}{\mean{r}} \ .
\end{equation}
In order to transform the stimulus parameters $a_i$, $b_i$, $c_i$ introduced in Eqs.\ (\ref{stim_params_1}) and (\ref{stim_params_2}), and to reveal dependences between them, we calculate the mean values
\begin{multline}
\mean{\left(\frac{r}{\tau}\right)^k e^{-r/\tau}}
= \left(-\frac{\partial}{\partial \alpha}\right)^k \mean{e^{-\alpha r/\tau}} \Bigg|_{\alpha=1}
= \left(-\frac{\partial}{\partial \alpha}\right)^k \int_0^{\infty} dr \, u(r) \, e^{-\alpha r/\tau} \Bigg|_{\alpha=1} \\
= \left(-\frac{\partial}{\partial \alpha}\right)^k \frac{1}{1 + \alpha \mean{r/\tau}} \Bigg|_{\alpha=1} \ .
\end{multline}
Hence the stimulus parameters turn out to be
\begin{equation}
a_1 = \frac{1}{1 + \mean{r/\tau}} \ , \quad b_1 = \frac{e \mean{r/\tau}}{\left(1 + \mean{r/\tau}\right)^2} \ ,
\end{equation}
\[
a_2 = \frac{1}{1 + 2 \mean{r/\tau}} \ , \quad b_2 = \frac{2 e \mean{r/\tau}}{\left(1 + 2 \mean{r/\tau}\right)^2} \ , \quad
c_2 = \frac{2 e^2 \mean{r/\tau}^2}{\left(1 + 2 \mean{r/\tau}\right)^3} \ .
\]
Dependences between these parameters are now explicit. In particular, we have
\begin{equation}
\label{Poisson_ai_bi_dep}
b_i = e \, a_i \, (1 - a_i) \ , \quad i = 1,2 \ .
\end{equation}

In Sec.\ \ref{var_V}, we have established that for Poisson statistics and small $\smean{r/\tau}$ the necessary and sufficient condition for the second moment of $V(t)$ to converge is $\gamma < 4/e$. Multiplying Eq.\ (\ref{Poisson_ai_bi_dep}) by $\gamma$, we see that this bound implies
\begin{equation}
\gamma b_i < 4 \, a_i \, (1 - a_i) < 4 \, a_i \quad\mbox{for $a_i \in (0,1)$, $i = 1,2$.}
\end{equation}
Conversely, $\gamma b_i < 4 a_i$ for all $a_i \in (0,1)$ together with Eq.\ (\ref{Poisson_ai_bi_dep}) implies $\gamma < 4/e$.

\end{document}